# On the Emission Pattern of Nanoscopic Emitters in Planar Anisotropic Matrix and Nanoantenna Structures†

**Pu Zhang**[1]

[1]School of Physics, Huazhong University of Science and Technology, Luoyu Road 1037, Wuhan, 430074, People's Republic of China

Email: puzhang0702@hust.edu.cn

**Peng-Long Ren**[1]

[1]School of Physics, Huazhong University of Science and Technology, Luoyu Road 1037, Wuhan, 430074, People's Republic of China

Email: penglong_ren@hust.edu.cn

**Xue-Wen Chen**[1]*

**\*Corresponding author**

[1]School of Physics, Huazhong University of Science and Technology, Luoyu Road 1037, Wuhan, 430074, People's Republic of China

Email: xuewen_chen@hust.edu.cn

Telephone: +86-027-8755 7374; +86-15107153502

Fax: +86-027-8755 6576

# Abstract

Single nanoscopic emitters embedded in crystalline matrix have become a valuable resource for emerging nanophotonic and quantum technologies. The generally anisotropic nature of the matrix strongly affects the emission properties of the quantum emitters, in particular, when the matrix is assembled in nanophotonic structures. We report on rigorous analysis and engineering of spontaneous emission from single emitters coupled to nanoantenna and planar anisotropic antenna structures. By developing a convenient theoretical method with efficient numerical implementation, we show that accurate modeling of the anisotropy is essential in predicting the emission pattern for many important systems, such as single molecules in solid-state matrix, isolated defects in 2D materials and so on. In particular, we illustrate the amplified effects of material anisotropy and geometrical anisotropy for emitters coupled to planar antenna and nanoantenna structures. We show that with proper design of the anisotropies strong enhancement of the emission rate and nearly collimated beam from single emitters can be simultaneously achieved.

**INTRODUCTION**

Single nanoscopic emitters embedded in solid-state matrix have been considered as one of the central building blocks in emerging nanophotonic and quantum technologies.[1-3] The solid-state platform promises scalability of the system size and controllability over the atom-photon interaction through chip integration and nano-optics techniques.[4-6] In order to obtain good photostability, minimum spectral jumps and spectral diffusion of the emitters, the solid-state matrix should be single crystalline to have a weak electron-phonon coupling and isolate the emitters from the noisy environment.[7,8] The single crystalline lattice structures usually imply strong optical anisotropy. Indeed, many important material systems, such as terrylene molecules in para-terphenyl,[9,10] dibenzoterrylene molecules in anthracene,[11,12] defects in two-dimensional (2D) hexagonal boron nitride (h-BN)[13-15], excitons in 2D molybdenum disulfide[16] and in perovskite stacks,[17,18] to name a few, fall into this category. Despite the fact that anisotropy leads to significant birefringence and substantially impacts the emission properties[19], there is little effort (even theoretical) in studying the effects due to anisotropy.

The solid-state realization facilitates the use of optical antenna structures to enhance the spontaneous emission rate and to direct the single-photon streams into well-defined optical modes.[20-28] The optical antennas usually have and enhance the anisotropic effects. For instance, nanoantennas based on plasmonic resonances of metallic nanoparticles have strong geometric anisotropic effect, greatly enhancing or quenching the spontaneous emission of the emitter depending on its relative orientation with the antenna.[21,29,30] One could expect that the material anisotropy of the matrix in the presence of the nanoantenna (geometric anisotropy) will be amplified. Other types of antennas, in particular, planar antenna structures relying on weak Fabry-Perot resonance or interference effect have been proposed to engineer the emission pattern and achieve up to near-unity collection efficiency.[31-34] However, due to resonant enhancement of the effect from the anisotropic matrix in the antenna, a direct migration of the design from the isotropic structure will not work.

In this work, we report on rigorous analysis and engineering of spontaneous emission pattern from single nano-emitters embedded in anisotropic matrix and planar antenna structures. Here the nano-emitter refers to a quantum emitter or a system consisting of an emitter coupled with a plasmonic nanoantenna. By developing a theoretical method and a convenient numerical implementation procedure, we show that accurate modeling of the anisotropy of the matrix in both planar antenna and nanoantenna systems is essential. Moreover, we demonstrate that with proper design of the anisotropy strong enhancement of the emission rate and collimated emission from a single nano-emitter can be simultaneously achieved.

**THEORY AND IMPLEMENTATION**

To set the ground for discussion, we consider a structure depicted in Fig. 1(a). A single dipole emitter and a nearby nanoantenna (e.g., a metallic nanoparticle with plasmon resonance) are embedded in a multilayer background structure. The background structure consists of $N$ layers of medium generally characterized by the permittivity tensors $\bar{\bar{\varepsilon}}_i$ and permeability tensors $\bar{\bar{\mu}}_i$ ($i = 1… N$) with symmetry property $\bar{\bar{\varepsilon}}_i^T = \bar{\bar{\varepsilon}}_i$ and $\bar{\bar{\mu}}_i^T = \bar{\bar{\mu}}_i$. The upper and lower half-spaces are assumed to be isotropic and lossless, and to be described by $\varepsilon_U$ and $\varepsilon_L$, respectively. Both the emission rate and emission pattern from the emitter are strongly dependent on the coupling of the emitter with the nanoantenna and the planar multilayer structure. The total emission rate and the dynamics of the emitter can be rigorously modelled by employing near-field calculations of the electromagnetic field and the analysis of emitter-field interaction.[35] In stark contrast, it is challenging to obtain the far-field emission pattern, since unlike in isotropic homogenous (or half-space) background[36-39] the dyadic Green's function in a general anisotropic multilayer structure is intricate to calculate.[40,41] An alternative approach based on Lorentz reciprocity can be used to compute the scattering pattern of a localized scatterer in the isotropic multilayer.[42,43] Here for the first time we extend the above idea to anisotropic system.

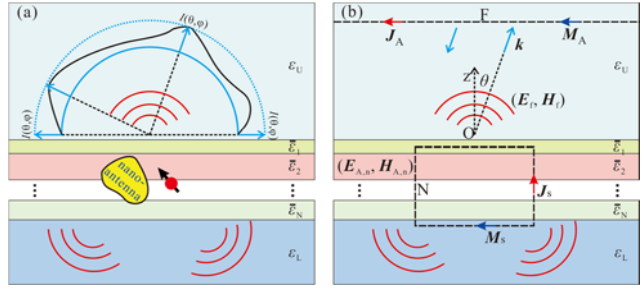

**Fig. 1** The schematic of the multilayer configuration under study consisting of *N* anisotropic layers and isotropic backgrounds. (a) The emission from dipolar quantum emitters is redirected by nearby nanoantennas and multilayers, and forms the radiation pattern *I(θ,φ)* in the far-field zone. (b) Local inhomogeneities including quantum emitters and nanoantennas are replaced with equivalent currents on an enclosing surface; Auxiliary surface currents are assumed on the horizontal plane denoted with dashed line.

In the following we briefly outline the theoretical method. To retrieve the emission pattern in the upper-most half-space, the electric field at the far field $\boldsymbol{E}_\text{f}$ can be expanded in terms of plane waves,

$$\boldsymbol{E}_\text{f} = \iint A_s(\hat{\boldsymbol{k}})(\hat{\boldsymbol{k}}_t \times \hat{z})e^{-i\boldsymbol{k}\cdot\boldsymbol{r}}dk_x dk_y + \iint A_p(\hat{\boldsymbol{k}})(\hat{\boldsymbol{k}}_t \times \hat{z}) \times \hat{\boldsymbol{k}} e^{-i\boldsymbol{k}\cdot\boldsymbol{r}}dk_x dk_y \quad (1)$$

where $\boldsymbol{k}_t$ is the in-plane component of the wave vector $\boldsymbol{k}$ and the hat stands for the unit vector. The time convention $e^{i\omega t}$ has been suppressed for simplicity. The main task is to find the amplitudes $A_{s,p}(\hat{\boldsymbol{k}})$ since the emission pattern can be obtained as $I(\theta,\phi) = \left(|A_s|^2 + |A_p|^2\right)\cos^2\theta/(2Z)$, where $Z$ is the wave impedance of the half-space and s or p denotes the state of polarization. The factor $\cos^2\theta$ accounts for the ratio between the area element in $\boldsymbol{k}$-space and its projection on $k_x$-$k_y$ plane. To apply the Lorentz reciprocity theorem, two sets of surface currents are concerned. Fig. 1(b) illustrates, (i) the first set, i.e. the equivalent surface currents $\boldsymbol{J}_S$ and $\boldsymbol{M}_S$ replacing the emitter and nanoantenna in the dashed rectangle, and (ii) the second set, i.e., a set of auxiliary surface currents $\boldsymbol{J}_A$ and $\boldsymbol{M}_A$ at the far field, which generate incident plane waves in $-\boldsymbol{k}'$ direction. The Lorentz reciprocity theorem reads[44]

$$\iint_N (\boldsymbol{E}_{A,n} \cdot \boldsymbol{J}_S - \boldsymbol{H}_{A,n} \cdot \boldsymbol{M}_S) dS = \iint_F (\boldsymbol{E}_f \cdot \boldsymbol{J}_A - \boldsymbol{H}_f \cdot \boldsymbol{M}_A) dS \qquad (2)$$

Here $\boldsymbol{E}_{A,n}$ and $\boldsymbol{H}_{A,n}$ are the fields at the near field due to $\boldsymbol{J}_A$ and $\boldsymbol{M}_A$. The integrals over N and F imply the surface area at the near field and far field as shown in Fig. 1(b), respectively. In view of the orthogonality of plane waves, the RHS of Eq.(2) allows analytical simplification and only the components with wave vector $\boldsymbol{k}'$ in the integral survive. By evaluating the LHS numerically, we obtain

$$A_{s,p}(\hat{\boldsymbol{k}}') = -\frac{Z}{8\pi^2 \cos\theta} \iint_N \left[ (\boldsymbol{H}_n \times \boldsymbol{E}_{A,n}) + (\boldsymbol{E}_n \times \boldsymbol{H}_{A,n}) \right] \cdot d\boldsymbol{S} \qquad (3)$$

$\boldsymbol{H}_n$ and $\boldsymbol{E}_n$ are the fields due to the dipole emitter. The detailed derivation is provided in Section I of the Supporting Information. The information of anisotropy hides in the field components appearing in Eq.(3). The auxiliary fields at the near-field ($\boldsymbol{E}_{A,n}$, $\boldsymbol{H}_{A,n}$) can be calculated by rewriting the Maxwell equations as $\partial \boldsymbol{\psi}/\partial z = -ik_0 M \boldsymbol{\psi}$.[45] Here the field vector is $\boldsymbol{\psi} = (E_x, E_y, Z_0 H_x, Z_0 H_y)^T$ in SI units and the anisotropic medium is characterized by a 4×4 matrix

$$M = \begin{bmatrix} -k_{tn}\epsilon_{zx}/\epsilon_{zz}, & -k_{tn}\epsilon_{zy}/\epsilon_{zz}, & 0, & 1 - k_{tn}^2/\epsilon_{zz} \\ 0, & 0, & -1, & 0 \\ \epsilon_{yz}\epsilon_{zx}/\epsilon_{zz} - \epsilon_{yx}, & k_{tn}^2 - \epsilon_{yy} + \epsilon_{yz}\epsilon_{zy}/\epsilon_{zz}, & 0, & k_{tn}\epsilon_{yz}/\epsilon_{zz} \\ \epsilon_{xx} - \epsilon_{xz}\epsilon_{zx}/\epsilon_{zz}, & \epsilon_{xy} - \epsilon_{xz}\epsilon_{zy}/\epsilon_{zz}, & 0, & -k_{tn}\epsilon_{xz}/\epsilon_{zz} \end{bmatrix} \qquad (4)$$

with $k_{tn} = k_t/k_0$. Each homogeneous layer then admits a transfer matrix linking the fields at adjacent interfaces $\boldsymbol{\psi}_j/\boldsymbol{\psi}_{j-1} = e^{-ik_0 M_j (z_j - z_{j-1})}$, and the auxiliary fields at any location can be calculated via standard transfer matrix method. The knowledge of $I(\theta, \phi)$ also helps determine the radiation efficiency $\eta = P_{rad}/P_{tot}$. The radiative power into the far field is calculated as[46]

$$P_{rad} = \iint (2\pi)^2 I(\theta,\phi) k^2 \sin\theta d\theta d\phi \tag{5}$$

and $P_{tot}$ can be obtained by integrating the Poynting vector in the near field over a surface enclosing the emitter.

The implementation of the above theoretical formulation essentially consists of three steps: (i) compute the near fields due to the dipole emitter with an electromagnetic solver; (ii) calculate the auxiliary fields at the near-field boundary; (iii) numerically evaluate the integral in Eq. (3). We have carefully validated the implementation with two examples. Plane wave scattering by a silicon sphere in vacuum is first considered as a test for the special case of isotropic medium. The good agreement in radiation pattern with the results reported in literature[43] and energy conservation show that the implementation works correctly for isotropic medium. A more stringent test concerns the radiation of an emitter embedded in an anisotropic slab, which was treated with dyadic Green's function technique.[40] Our implementation can conveniently reproduce the radiation patterns, and thus it is reliable. The details are given in Section 2 of the Supporting Information.

**RESULTS AND DISCUSSION**

In this section, we investigate the emission properties of several important solid-state single quantum systems with coupling antenna structures. Firstly, we study the emission pattern of the emitter in the anisotropic matrix assembled to a planar antenna structure. Then we analyze the anisotropic emission properties of an emitter in the anisotropic matrix coupled to a plasmonic nanoantenna. In the end, we study and engineer the synergistic effect of anisotropic planar antenna and nanoantenna by coupling them with the emitter embedded in anisotropic matrix. We will show that it is possible to achieve simultaneously strong enhancement of the emission rate and collimated emission to achieve a high collection efficiency.

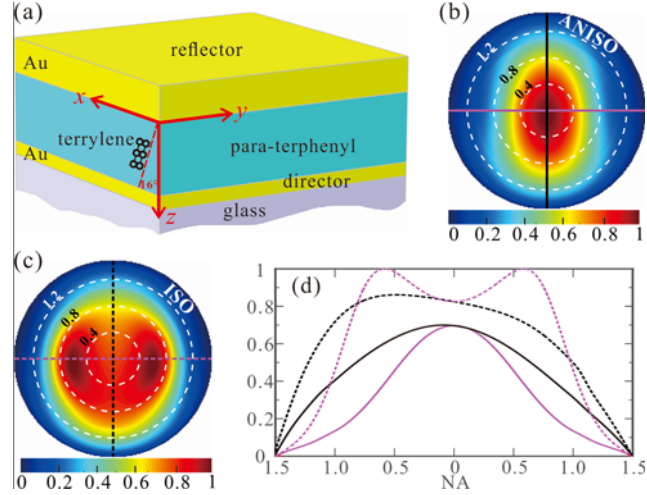

**Fig. 2** (a) A single terrylene molecule is embedded at the center of a 100 nm para-terphenyl layer sandwiched between two gold layers (100 nm and 20 nm); the emission dipole parallels x-z plane with $\theta$=16°. The emission patterns from the single terrylene molecule are displayed in the form of back focal plane (BFP) images when the para-terphenyl matrix is treated as anisotropic (b) and isotropic (c). (d) Intensity distributions along the cross sections in (b) and (c) are plotted with corresponding line styles.

We first investigate the case of single terrylene molecules in para-terphenyl matrix, which has been studied as a model system since early days of single-molecule spectroscopy.[47,48] It has long been known that the para-terphenyl crystal shows biaxial anisotropy with principal refractive indices of $n_x$=1.59, $n_y$=1.69 and $n_z$=2.0. However, the effects of the anisotropic matrix on the emission pattern of the terrylene molecule have not been carefully studied yet. We explore the possibility of directing emission of this system into a narrow angle by introducing a planar nanoantenna and study the role of anisotropy. The configuration comprises a crystalline para-terphenyl film sandwiched between two gold layers as shown in Fig. 2(a) like a planar Yagi-Uda antenna.[26,33,49] The reflector and director layers are 100 nm and 20 nm thick, respectively. The terrylene molecules locate at the center of the matrix and radiate at the wavelength of 580 nm. The emission dipole of the emitters parallels to the x-z plane with a tilting angle of $\theta = 16°$.[10] We studied the emission pattern for the system as a function of the matrix thickness and found that photon beaming effect can be obtained for a matrix with a thickness of 100 nm. The emission pattern corresponding to the image measurable at the back focal plane (BFP) of an objective is illustrated in Fig. 2(b). We observe that the emission is directed into an elliptical-shaped area in k space and 80% of the emission can be collected by an objective with NA=1.22. In comparison, we also approximated the para-terphenyl matrix as isotropic medium with a refractive index of 1.85 as often treated in literature.[10] Fig. 2(c) displays the emission pattern for the isotropic case. One sees that the pattern is completely different from the anisotropic case and there are two side lobes instead of only a central one. Fig. 2(d) further depicts the distributions along the four cross sections denoted in the BFP images with different line styles. The isotropic approximation (dashed curves) apparently leads to wider angular distribution in both cross sections. The black curves show peaks shifted from the center. For the isotropic case, the asymmetry is the result of the tilting emission dipole. The asymmetry is largely counteracted in the anisotropic case by the anisotropy of the solid matrix. We remark that the resonance effect of the planar antenna structure amplifies the anisotropy of para-terphenyl matrix and dramatically changes the emission pattern.

The second case concerns the emission from isolated defects in h-BN matrix, which is a promising candidate for realizing bright single-photon sources.[2,13] As one of the emerging 2D materials,[50] h-BN is well known to be optically anisotropic[51-53] and yet there is little theoretical study towards the effect of anisotropy. Here we consider a single defect emitter embedded in the anisotropic matrix and coupled to a nanoantenna. Specifically, the system consists of a 100 nm thick h-BN flake on a glass ($n_g$=1.5) coverslip and a nanoantenna as shown in Fig. 3(a). The optic axis of the h-BN matrix is in z direction, and the refractive indices are $n_z$=2.13 and $n_\perp$=1.65, respectively, according to experimental measurements.[53] Previous works have shown that the defect emitters have in-plane dipole moments and the emission spectra have narrow peaks covering visible and near infrared spectral range depending on the experimental conditions.[13] Here we considered the emitters with an emission peak

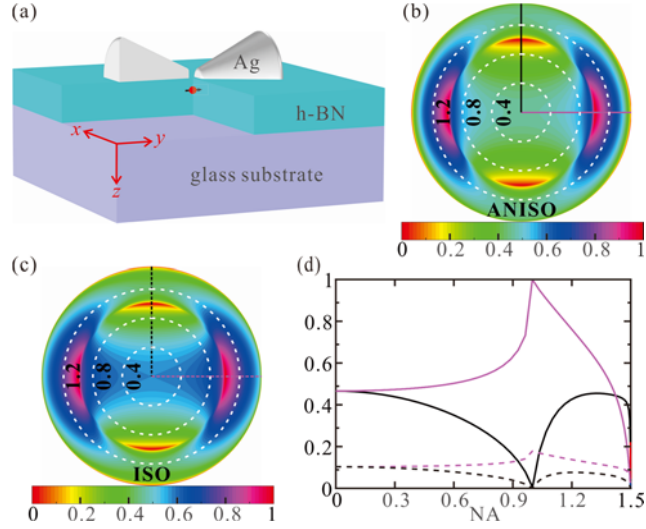

**Fig. 3** (a) A defect emitter embedded inside a 100 nm thick h-BN flake locates 10 nm below the top surface; the emission is enhanced by a silver half-cone dimer nanoantenna with the axis aligned with the emission dipole. A quarter of the nanoantenna and h-BN matrix is hidden to better visualize the whole structure. (b) and (c) display the emission patterns when h-BN is modeled as anisotropic and isotropic, respectively. (d) Intensity distributions along the cross sections in (b) and (c).

around 650 nm and designed the nanoantenna with a matching plasmonic resonance. We assume that the emission dipole orients in y direction and situates 10 nm below the top surface. The nanoantenna consists of two silver half cones with a 5 nm gap. The cones have a tip radius of 5 nm (flat), a base radius of 50 nm and a height of 40 nm. The spectral response of the nanoantenna is shown in Fig. S3 of the Supporting Information. The emission pattern of the defect emitter coupled to the nanoantenna is rendered in the form of BFP image as in Fig. 3(b). For comparison, we also calculated the emission pattern for the isotropic approximation of the h-BN matrix with a refractive index $n_{BN} = 1.8$ (averaged value of the indices of three axes) and displayed the resulting pattern in Fig. 3(c). We observe that the central parts of the two patterns are distinct. The isotropic approximation incorrectly overestimates the intensity distribution in the center. Even larger discrepancies are disclosed when the absolute values of the emission patterns are visualized as in Fig. 3(d), where the distributions along the four cross sections in the BFP images are plotted in correspondingly styled curves. Anisotropy manifests itself in another indirect but dramatic way. Isotropic approximation fails to capture the resonance of the nanoantenna, leading to inferior emission performance with the Purcell factor $f_P^I$ =157 and quantum efficiency $\eta^I$ =53.2%. Lifting the approximation, the emission is greatly boosted by the nanoantenna with $f_P^A$=775 and quantum efficiency $\eta^A$=57.3%. For completeness, we also examined the isotropic approximation with $n_{BN}$ = 1.65. The results included in Section 4 of the Supporting Information show that it is a generally worse approximation as the index deviates more from the anisotropic refractive index in the average sense.

In the third example, we combine the planar antenna and nanoantenna design to simultaneously achieve strong enhancement of the radiative decay rate and collimated emission. As shown in Fig. 4(a), we sandwich the defect emitter and nanoantenna between a silver reflector and director of 100 nm and 20 nm thicknesses, respectively. Instead of full optimization, we only adjusted the thickness of the air layer to tailor the emission pattern and decay rate. The composite system is found to exhibit remarkable beaming effect with a 60 nm thick air layer as shown in Fig. 4(b). The distributions along the cross sections denoted by the solid lines in the BFP image are depicted in Fig. 4(d), and illustrate that a dominant majority of photons can be collected within small angles, specifically, 90% photons within 29.1° or NA=0.729, and 80% within 23.6° or NA=0.601. In contrast, under isotropic approximation, as shown in Fig. 4(c) and 4(d) (dashed lines), no beaming effect is predicted and the radiative emission rate remains much lower than the anisotropic case. With accurate modeling of the anisotropy, we find the dimer nanoantenna still performs rather well with $f_P^A = 620$ and $\eta = 29.3\%$. Here the reduced radiation efficiency is mainly due to the coupling of the emission to the in-plane waveguide modes while the absorption in the plasmonic nanoantenna is basically unchanged. We believe higher radiative efficiency is possible by allowing more freedom in the optimization.

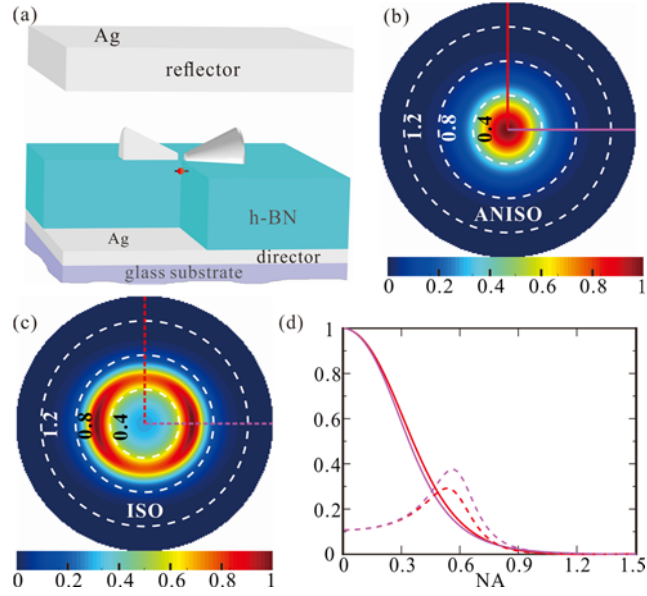

**Fig. 4** (a) A single defect emitter in h-BN matrix coupled to a half-cone dimer nanoantenna (*c.f.* Fig. 3(a)) is sandwiched between a silver planar antenna; the reflector and director are 100 nm and 20 nm thick, respectively. (b) The emission pattern of the emitter when h-BN is correctly modelled. (c) The emission pattern when h-BN is modelled as an isotropic medium. (d) Intensity distributions along the cross sections in (b) and (c).

## CONCLUSIONS

In this work, we have revealed the impact of anisotropic matrix on the emission properties of single quantum emitter, which can be decisive especially when coupled to planar antenna structures, nanoantennas, or both. We have developed a generic theoretical and numerical method for the analysis of the emission pattern of emitters in anisotropic medium. Taking experimentally important systems as examples, anisotropy is found to play crucial roles through two mechanisms: the anisotropic effect is either directly amplified by resonant antennas, or manifests itself indirectly but strongly by affecting the resonance of nanoantennas. We have shown that, combining dimer and planar nanoantennas, greatly enhanced emission rate and highly-directional beaming can be accomplished simultaneously. Specifically, a Purcell factor of 620 and collection of 90% radiated photons within $\theta = 30°$ are numerically demonstrated with experimentally realizable parameters. These findings might be important for photonic quantum technologies based on solid-state single quantum systems.

## AUTHOR INFORMATION

**Corresponding Author**

*E-mail: xuewen_chen@hust.edu.cn.

**Notes**

The authors declare no competing financial interest.

## ACKNOWLEDGEMENTS

We acknowledge financial support from the National Natural Science Foundation of China (grant 11474114, 11604109, 11874166), the Thousand-Young-Talent Program of China, Huazhong University of Science and Technology.

## REFERENCES

(1) P. Lodahl and S. Stobbe, *Nanophotonics*, 2013, **2**, 39-55.
(2) I. Aharonovich, D. Englund and M. Toth, *Nat.* Photonics, 2016, **10**, 631-641.
(3) D. D. Awschalom, R. Hanson, J. Wrachtrup and B. B. Zhou, *Nat. Photonics*, 2018, **12**, 516-527.
(4) J. L. O'Brien, A. Furusawa and J. Vuckovic, *Nat. Photonics*, 2009, **3**, 687-695.
(5) O. Benson, *Nature*, 2011, **480**, 193-199.
(6) K. G. Lee, X. -W. Chen, H. Eghlidi, P. Kukura, R. Lettow, A. Renn, V. Sandoghdar and S. A. Götzinger, *Nat. Photonics*, 2011, **5**, 166-169.
(7) T. Basché, W. E. Moerner, M. Orrit and U. P. Wild, U. P. *Single-Molecule Optical Detection, Imaging and Spectroscopy*, VCH Verlagsgesellschaft mbH, 2007.
(8) C. Gooijer, F. Ariese and J. W. Hofstraat, *Shpol'Skii Spectroscopy and Other Site Selection Methods*, John Wiley & Sons, 2000.
(9) B. Lounis and W. E. Moerner, *Nature*, 2000, **407**, 491-493.
(10) B. C. Buchler, T. Kalkbrenner, C. Hettich and V. Sandoghdar, *Phys. Rev. Lett*., 2005, **95**, 063003-063006.
(11) A. A. Nicolet, P. Bordat, C. Hofmann, M. A. Kol'chenko, B. Kozankiewicz, R. Brown and M. Orrit, *ChemPhysChem*, 2007, **8**, 1929-1936.

(12) P. Lombardi, A. Ovvyan, S. Pazzagli, G. Mazzamuto, G. Kewes, O. Neitzke, N. Gruhler, O. Benson, W. Pernice and F. Cataliotti, *ACS Photonics*, 2018, **5**, 126-132.
(13) T. T. Tran, C. Elbadawi, D. Totonjian, C. J. Lobo, G. Grosso, H. Moon, D. R. Englund, M. J. Ford, I. Aharonovich and M. Toth, *ACS Nano*, 2016, **10**, 7331-7338.
(14) M. Nguyen, S. Kim, T. T. Tran, Z. Q. Xu, M. Kianinia, M. Toth and I. Aharonovich, *Nanoscale*, 2018, **10**, 2267-2274.
(15) A. L. Exarhos, D. A. Hopper, R. R. Grote, A. Alkauskas and L. C. Bassett, *ACS Nano*, 2017, **11**, 3328-3336.
(16) D. B. Hu, X. X. Yang, C. Li, R. N. Liu, Z. H. Yao, H. Hu, S. N. G. Corder, J. N. Chen, Z. P. Sun, M. K. Liu and Q. Dai, *Nat. Comm.*, 2017, **8**, 1471-1478.
(17) Y. S. Park, S. Guo, N. S. Makarov and V. I. Klimov, *ACS Nano*, 2015, **9**, 10386-10393.
(18) N. Zhang, K. Wang, H. Wei, Z. Gu, W. Sun, J. Li, S. Xiao and Q. Song, *J. Phys. Chem. Lett.*, 2016, **7**, 3886-3891.
(19) I. J. Hodgkinson and Q. H. Wu, *Birefringent Thin Films and Polarizing Elements*, World Scientific Publishing Co. Pte. Ltd.: Singapore, 1997.
(20) J. J. Greffet, *Science*, 2005, **308**, 1561-1563.
(21) L. Rogobete, F. Kaminski, M. Agio and V. Sandoghdar, *Opt. Lett.*, 2007, **32**, 1623-1625.
(22) R. Esteban, T. V. Teperik and J. J. Greffet, *Phys. Rev. Lett.*, 2010, **104**, 026802-026805.
(23) G. Lozano, D. J. Louwers, S. R. K. Rodriguez, S. Murai, O. T. Jansen, M. A. Verschuuren and J. Gomez Rivas, *Light: Sci. Appl.*, 2013, **2**, e66.
(24) N. Gregersen, T. R. Nielsen, J. Claudon, J. M. Gérard and J. Mørk, *Opt. Lett.*, 2008, **33**, 1693-1695.
(25) G. C. Li, Y. L. Zhang and D. Y. Lei, *Nanoscale*, 2016, **8**, 7119-7126.
(26) A. G. Curto, G. Volpe, T. H. Taminiau, M. P. Kreuzer, R. Quidant and N. F. van Hulst, *Science*, 2010, **329**, 930-933.
(27) L. Novotny and N. F. van Hulst, *Nat. Photonics*, 2011, **5**, 83-90.
(28) M. Agio, *Nanoscale*, 2012, **4**, 692-706.
(29) Y. Gu, L. J. Wang, P. Ren, J. X. Zhang, T. C. Zhang, O. J. F. Martin and Q. Gong, *Nano Lett.*, 2012, **12**, 2488-2493.
(30) B. Gurlek, V. Sandoghdar and D. Martín-Cano, *ACS Photonics*, 2018, **5**, 456-461.
(31) K. G. Lee, X.-W. Chen, H. Eghlidi, P. Kukura, R. Lettow, A. Renn, V. Sandoghdar and S. A. Götzinger, *Nat. Photonics*, 2011, **5**, 166-169.
(32) X.-W. Chen, S. A. Götzinger and V. Sandoghdar, *Opt. Lett.*, 2011, **36**, 3545-3547.
(33) S. Checcucci, P. Lombardi, S. Rizvi, F. Sgrignuoli, N. Gruhler, F. B. Dieleman, F. S. Cataliotti, W. H. Pernice, M. Agio and C. Toninelli, *Light: Sci. Appl.*, 2016, **6**, e16245.
(34) A. Pors and S. I. Bozhevolnyi, *ACS Photonics*, 2015, **2**, 228-236.
(35) X.-W. Chen, V. Sandoghdar and M. Agio, *Phys. Rev. Lett.*, 2013, **110**, 153605-

153609.

(36) T. K. Sarkar and A. Taaghol, *IEEE Trans. Antenn. Propag.*, 1999, **47**, 566-573.

(37) J. Muller, G. Parent, G. Jeandel and D. Lacroix, *J. Opt. Soc. Am. A*, 2011, **28**, 868-878.

(38) A. Taflove and S. C. Hagness, *Computational Electrodynamics: The Finite-Difference Time-Domain Method*, 3rd ed., Artech: Boston, 2005.

(39) J. M. Jin, *The Finite Element Method in Electromagnetics*, 2nd ed., Wiley-IEEE Press, 2002.

(40) A. Eroglu and J. K. Lee, *IEEE Trans. Antennas Propag.*, 2005, **53**, 3963-3973.

(41) L. Pazynin, S. Sautbekov, Y. Sirenko, A. Vertiy and N. Yashina, *Telecommun. Radio Eng.*, 2015, **74**, 1039-1050.

(42) K. Demarest, Z. Huang and R. Plumb, *IEEE Trans. Antennas Propag.*, 1996, **44**, 1150-1157.

(43) J. Yang, J. P. Hugonin and P. Lalanne, *ACS Photonics*, 2016, **3**, 395-402.

(44) R. F. Harrington, *Time Harmonic Electromagnetic Fields*, John Wiley & Sons, 2001.

(45) D. W. Berreman, *J. Opt. Soc. Am.*, 1972, **62**, 502-510.

(46) R. C. Johnson, H. A. Ecker and J. S. Hollis, *Proc. IEEE*, 1973, **61**, 1668-1694.

(47) W. E. Moerner, T. Plakhotnik, T. Irngartinger, M. Croci, V. Palm and U. P. Wild, *J. Phys. Chem.*, 1994, **98**, 7382-7389.

(48) S. Kummer, T. Basche and C. Brauchle, *Chem. Phys. Lett.*, 1994, **229**, 309-316.

(49) T. Kosako, Y. Kadoya and H. F. Hofmann, *Nat. Photonics*, 2010, **4**, 312-315.

(50) K. Novoselov, A. Mishchenko, A. Carvalho and A. C. Neto, *Science*, 2016, **353**, aac9439.

(51) R. Geick, C. H. Perry and G. Rupprecht, *Phys. Rev.*, 1966, **146**, 543-547.

(52) V. Bougrov, M. E. Levinshtein, S. L. Rumyantsev and A. Zubrilov, *Properties of Advanced Semiconductor Materials: GaN, AlN, InN, BN, SiC, SiGe*, John Wiley & Sons, Inc.: New York, 2001.

(53) T. Ishii and T. Sato, *J. Cryst. Growth*, 1983, **61**, 689-690.